\def\cn{C$_{3}$N$_{4}$}
\def\etal{{\it et al. }}

 \documentclass[final,3p,times,twocolumn]{elsarticle}
\usepackage{amssymb}
\usepackage{amsmath}
\usepackage{float}
\usepackage{subfigure}
\usepackage{color}
\usepackage{adjustbox}
\usepackage{color, colortbl}
\definecolor{Gray}{gray}{0.75}
\definecolor{LightCyan}{rgb}{0.50,1,1}

\usepackage[section]{placeins}
\usepackage{amsfonts}
\usepackage{graphicx}
\usepackage{lipsum}
\usepackage{multirow}
\usepackage{booktabs}
\usepackage[table,xcdraw]{xcolor}
\usepackage{ulem}
\definecolor{cream}{RGB}{222,217,201}

\def\vr{{\bf r}}

\journal{Carbon}
\begin {document}
\begin{frontmatter}

\title{Exploring the role of electronic structure on photo-catalytic behavior of carbon-nitride polymorphs}

\author{Sujoy Datta}
\address{Department of Physics, University of Calcutta, Kolkata 700009, India}
\address{Department of Physics, Lady Brabourne College, Kolkata 700017, India}
\author{Prashant Singh \fnref{fn1}}
\fntext[fn1]{Corresponding author: Tel: +1-515-294-2457/2122. \\ Email: psingh84@ameslab.gov, ddj@ameslab.gov}
\address{Ames Laboratory, U.S. Department of Energy, Iowa State University,  Ames, Iowa 50011, USA}
\author{Debnarayan Jana}
\address{Department of Physics, University of Calcutta, Kolkata 700009, India}
\author{Chhanda B Chaudhuri} 
\address{Department of Physics, Lady Brabourne College, Kolkata 700017, India}
\author{Manoj K Harbola} 
\address{Department of Physics, Indian Institute of Technology, Kanpur, 208016, India}
\author{Duane D. Johnson \fnref{fn1}} 
\address{Ames Laboratory, U.S. Department of Energy, Iowa State University,  Ames, Iowa 50011, USA}
\address{Materials Science \& Engineering, Iowa State University, Ames, Iowa 50011, USA}
\author{Abhijit Mookerjee} 
\address{S. N. Bose National Centre for Basic Sciences, Salt Lake City, Kolkata 700098, India}

\begin{abstract} 
A fully self-consistent density-functional theory (DFT) with improved functionals is used to provide a comprehensive account of structural, electronic, and optical properties of \cn~polymorphs. Using our recently developed van Leeuwen-Baerends (vLB) corrected local-density approximation (LDA), we implemented LDA+vLB within full-potential N$^{th}$-order muffin-tin orbital (FP-NMTO) method and show that it improves structural properties and band gaps compared to semi-local functionals (LDA/GGA).  We demonstrate that the LDA+vLB predicts band-structure and work-function for well-studied 2D-graphene and bulk-Si in very good agreement with experiments, and more exact hybrid functional (HSE) calculations as implemented in the Quantum-Espresso (QE) package.  The structural and electronic-structure (band gap) properties of \cn~polymorphs  calculated using FP-NMTO-LDA+vLB is compared with more sophisticated hybrid-functional calculations. We also perform detailed investigation of photocatalytic behavior using QE-HSE method of \cn~polymorphs through work-function, band (valence and conduction) position with respect to water reduction and oxidation potential. Our results show $\gamma$-\cn~as the best candidate for photocatalysis among all the \cn~polymorphs but it is dynamically unstable at `zero' pressure. We show that $\gamma$-\cn~can be stabilized under hydrostatic-pressure, which improves its photocatalytic behavior relative to water reduction and oxidation potentials. 
\end{abstract}

\begin{keyword}Density-functional theory \sep graphene \sep semiconductor  \sep band gap \sep photocatalysis

\end{keyword}

\end{frontmatter}

\section{Introduction}

Conjugated carbon-nitride (C$_{3}$N$_{4}$) polymers have drawn broad interdisciplinary attention as metal-free and visible-light-responsive photo-catalysts for solar-energy conversion and environmental remediation \cite{Wang2009,Wang2009_1,2_2},  especially for their appealing electronic dispersion, high physicochemical stabilities, and earth-abundant elements.
{Though experimental works on graphitic carbon nitrides, both on triazine (t-g) and heptazine (h-g) structures, are prolific, the low specific surface area and high defect density remains a concern on its photocatalytic performance. Among other experimental techniques, combination of graphitic-\cn with three dimensional hydrogel framework, exfoliation into nanosheets, construction of graphitic-\cn~based heterojunction systems are widely investigated \cite{hu2019recent, fu2018g}. Beside experimental trials, a plethora of first principles calculations are also aimed to improve the photocatalytic activity of graphitic-\cn~ and its derivatives \cite{zhu2018first, ma2012strategy}.} Notably, doping (e.g., by S or Si) or hybridization with other photocatalyst enhances the photo-reactive performance of graphitic-\cn~\cite{gcn,shayeganfar2016}. The large band gap tunability makes \cn~derivatives potential candidates for CO$_{2}$ capture, control of pollutants \cite{5_2}, water splitting, or energy-storage devices \cite{5_2,6_2}.

{\par} Carbon-nitride is found in amorphous and various crystalline forms \cite{Hemley1996,Guo1995}, with distinctly different characteristics (similar to two crystalline forms of carbon: graphite and diamond). However, first-principles electronic-structure and optical properties of the known \cn~polymorphs are sparse. Using density-functional theory (DFT) methods, we investigate the structural, electronic, and optical properties of \cn~polymorphs, specifically: (i)  $\alpha$, (ii) $\beta$, (iii) $cubic$, (iv) $\gamma$ (spinel), (v) $t-g$(AA) [AA-stacked graphitic-triazine], (vi)  $t-g$(AB) [AB-stacked graphitic-triazine], and (vii) $h-g$ [AA-stacked graphitic-heptazine]. 

{\par} The LDA or GGA exchange-correlation (XC) functionals \cite{PBE,PBEsol} are known to provide a reasonable account of structural properties, but show limitations and often miserably fail in predicting band gaps \cite{Mattesini2000,yao,Molina1999}. To address \cn~polymorphs, we use our recently developed van Leeuwen-Baerends (vLB) corrected LDA (LDA+vLB) exchange~\cite{PS2016}  implemented both in tight-binding linearized muffin-tin orbital {(TB-LMTO)}~\cite{VLB, TBLMTO} and in full-potential N$^{th}$-order muffin-tin orbital {(FP-NMTO)}~\cite{NMTO1}. The LDA+vLB is similar to a modified-LDA~\cite{PhysRevLett.102.226401} and mimics the orbital-dependent, exact-exchange potentials~\cite{SD2019}  and notably requires a computational cost similar to the semi-local XC, like LDA or GGA. {Notably, the advantage of our approach lies in the fact that the LDA+vLB greatly improves the band gap relative to traditional (semi)local functionals, with speed of semi-local functionals and accuracy of advanced hybrid-functional (e.g. HSE) or GW functionals \cite{PS2013,PS2015book,PS2016,PS2017,PhysRevB.96.054203,SD2019}}. Our LDA+vLB calculated band-structure of 2-dimensional (2D) graphene and bulk-Si show good agreement to experiments \cite{Si_expt} (see supplement), and with hybrid-functionals (HSE) \cite{HSEtheo} as implemented in the Quantum-Espresso (QE) package~\cite{QE1,QE2}). 

{\par} Considering the fact that most semi-local functionals underestimate band-energies and band gap \cite{Mattesini2000,Molina1999}, we employ self-consistent QE-HSE to calculate optical properties, work functions, and photocatalytic behavior of \cn~polymorphs. We calculate work-functions of well-studied 2D-graphene \cite{ziegler2011} and bulk-Si \cite{hollinger1983} to showcase agreement between prediction and experiments. Our study of \cn~polymorphs finds better optical and photocatalytic behavior of $\gamma$-\cn~compared to other polymorphs. However, we find $\gamma$-phase exhibits a dynamical instability, i.e., unstable phonon modes with imaginary frequencies. Therefore, we theoretically assess the structural stability versus pressure to find the dynamical stability of $\gamma$-phase, and then fully detail the unstable and stable phonon dispersion. Under pressure, the dynamically stable $\gamma$-\cn~ also satisfies the Born  structural stability criteria for the elastic constants of the structure. Our notable finding is that the dynamically stable $\gamma$-\cn~(under applied hydrostatic pressure) also shows much improved photocatalytic behavior. 

{In this systematic study, we established the robustness of our methodology and then we explored the relevant properties of~\cn~polymorphs such as band gap, optical properties, and photocatalytic behavior. Our study reveals the possibility of $\gamma$ phase as better photocalytic  material than graphitic phases, which are traditionally found to be a suitable photo-catalytic material. Beside this, the scheme can further be applied to the appropriate derivatives of~\cn~polymorphs and other emergent semiconducting materials.}

\section{Computational Method}
{\textbf{ FP-NMTO electronic-structure}:~} Choice of a minimal basis set is always tricky, but tight-binding and  FP-NMTO \cite{NMTO1} methods provide that platform. The FP-NMTO handles the orbital ($l$-dependent) and $m$-dependent downfoldings independently, which is very useful for $sp^{2}$-hybridized systems, where $p_{z}$ orbitals behaves differently than $p_{x}$ and $p_{y}$ \cite{NMTO1,NMTO2}. As most semi-local functionals fail to estimate the correct band gap of semiconducting materials, we predict band gaps of \cn~polymoprhs using an {\it {in-house}} developed LDA+vLB exchange with von-Barth-Hedin correlation for solids \cite{PS2013,PS2016}, as implemented in {FP-NMTO} \cite{NMTO1}. The LDA+vLB potential for given atomic densities ($\rho$) can be written as
\begin{eqnarray}
v_{xc}(\vr)= [v_{x}^{LDA}(\vr)+ v_{x}^{vLB}(\vr)]+v_{c}^{LDA}(\vr) ,  
\label{eq6}
\end{eqnarray}
\begin{eqnarray}
\text{with} \,\,\,    v_{x}^{vLB}(\vr)= -\beta \rho^{1/3}(\vr) \dfrac{z^2}{1+3 \beta\ z \sinh^{-1} (z)}
\label{eq7}
\end{eqnarray}
Here, $z={|{\nabla} \rho(\vr) |} /{\rho ^{4/3}(\vr)}$ and $\beta = 0.05$.

{\par} The LDA+vLB (here onwards we use LDA+vLB instead of FP-NMTO-LDA+vLB) provides improvement to the exchange-potential \cite{PS2017} and better band gaps than other semi-local functionals as gradient correction naturally provides the self-interaction correction (SIC) to the LDA without need of SIC-LDA \cite{kraisler2014fundamental}. Most semi-local functionals fail due to the wrong asymptotic  behavior at r$\rightarrow{0}$ and r$\rightarrow{\infty}$ limits, whereas LDA+vLB produces  asymptotically-correct, Coulomb-like ($-1/r$)  behavior \cite{Almb,PS2016}. 

{\par} The LDA+vLB energies are self-consistently converged to 10$^{-6}$~Ry/atom. We use Anderson method \cite{Amix1965} to mix charge densities. For k-space integration via the tetrahedron method, we use $k$-mesh of ($6\times6\times6$) for $\alpha$; ($6\times6\times10$) for $\beta$; ($4\times4\times4$) for both cubic and spinel ($\gamma$); ($6\times6\times8$) for t-g (AA); and ($6\times6\times4$) for t-g(AB) phases. 

{\par} {\textbf{QE-HSE Optical \& Work-Function Properties}:~} We performed band-structure calculations of \cn~polymorphs using HSE as implemented in the QE plane-wave basis \cite{QE1,QE2} and compare that with our LDA+vLB predictions (also see supplement). The LDA+vLB predicted band gaps show very good agreement (see Results) with hybrid-functional and experiments. In addition, the self-consistent QE-HSE is used to calculate the complex dielectric tensor $\epsilon_{\alpha\beta}(\omega)$ within the random phase approximation to analyze the work function and photocatalytic behavior of \cn~polymorphs \cite{QE1,QE2}. For details see supplement (Table S1).

{\par} The dielectric tensor $\epsilon_{\alpha\beta}(\omega)$ can be defined as:
\begin{eqnarray}
\epsilon_{\alpha\beta}(\omega)&= 1+\frac{4 \pi e^2}{\Omega N_{\textbf{k}} m^2}\sum\limits_{n,n'}\sum\limits_{\textbf{k}}
\frac{\langle u_{\textbf{k},n'}\vert\hat{\textbf{p}}_{\alpha}\vert u_{\textbf{k},n}\rangle 
	\langle u_{\textbf{k},n}\vert\hat{\textbf{p}}_{\beta}^{\dagger} \vert u_{\textbf{k},n'}\rangle}
{(E_{\textbf{k},n'}-E_{\textbf{k},n})^2} \nonumber\\
&\left[\frac{f(E_{\textbf{k},n})}{E_{\textbf{k},n'}-E_{\textbf{k},n}+\hbar\omega+i\hbar\Gamma} +
\frac{f(E_{\textbf{k},n})}{E_{\textbf{k},n'}-E_{\textbf{k},n}-\hbar\omega-i\hbar\Gamma}\right]    ,
\end{eqnarray}
where $\Gamma$ is an adiabatic  (inter-broadening) parameter tending to zero. To retain a finite lifetime of all excited-states,  we have introduced small positive value of $\Gamma$  to produce an intrinsic broadening to all exited states. The imaginary part of $\epsilon^i_{\alpha\beta}$ is found first and real part $\epsilon^r_{\alpha\beta}$ is calculated using Kramers-Kronig relation. 
$$\epsilon^r_{\alpha \beta}(\omega)=1+\frac{2}{\pi}\int_{0}^{\infty}\frac{\omega' \epsilon^i_{\alpha \beta}(\omega')} {\omega'^{2}-\omega^{2}}d\omega'$$.
Real and imaginary parts of the dielectric function are used to calculate optical conductivity, refractive index and absorption coefficient.\cite{wooten}
\begin{eqnarray}
\text{Dielectric tensor:        } \quad\quad\quad\,\,\,  \epsilon_{\alpha\beta}=\epsilon_{\alpha\beta}^r+ i \epsilon^i_{\alpha\beta} \\ 
\text{Optical Conductivity:  } Re [\sigma_{\alpha\beta} (\omega)]= \frac{\omega}{4\pi}\epsilon^i_{\alpha\beta}(\omega)
\end{eqnarray}
%\begin{flushleft}
\begin{eqnarray}
&\text{Refractive Index:  }\quad   \mu_{\alpha\alpha}=n_{\alpha\alpha}+i k_{\alpha\alpha}\\ \nonumber
%\end{eqnarray}
%\begin{eqnarray}
&  n_{\alpha\alpha}(\omega)= \sqrt{\frac{|\epsilon_{\alpha\alpha}|+\epsilon^r_{\alpha\alpha}}{2}};
\, \, \, k_{\alpha\alpha}(\omega)= \sqrt{\frac{|\epsilon_{\alpha\alpha}|-\epsilon^r_{\alpha\alpha}}{2}} \nonumber
\end{eqnarray}
%\end{flushleft}
\begin{equation}
\text{Absorption Coefficient: } A_{\alpha\alpha}(\omega)=\frac{2\omega k_{\alpha\alpha}(\omega)}{c}
\end{equation}

{\textbf {{Work-function calculation:}~}}{For photocatalytic activity, the valence-band maximum (VBM) and conduction-band minimum (CBM ) position are equally important as band gaps. Therefore, the accurate calculation of work-function ($\Phi$) becomes important, which is defined as the  energy required to remove an electron from a slab surface. The $\Phi$ is calculated by taking the difference of vacuum energy ($E_{vac}$) with respect to Fermi-energy (E$_{F}$), i.e., $\Phi=V_{vac}-E_{F}$ \cite{feng2016}. For semi-infinite surfaces (with sufficient number of layers that reproduces bulk behavior), $\Phi$ is the difference of vacuum energy for slab and the Fermi energy of the bulk ($\Phi = V_{vac}-E_F^{Bulk}$). This needs two different calculations, (a) bulk, and (b) slabs (semi-infinite surface). The electrostatic potential within the interstitial region in (a) and (b) works as a common parameter, which should be same for (a) bulk and (b) slab in order to compare the energy levels. The matching bulk and slab values standardizes the energy levels. Also, the presence of microscopic fluctuations of electrostatic potentials in the interstitial region require macroscopic averaging of average planar potential with a window size equal to the inter-slab separation. The average planar potential is the average over the plane parallel to the surface-plane and the direction of microscopic averaging  (of planar potential) is perpendicular to that plane \cite{macro_av,macro_av2}. Here, the difference of the macroscopic averages for bulk and slab ($\Delta_V = V^{bulk} - V^{slab}$) works as a constant correctional shift to $E_F^{Bulk}$ that brings equivalence of energy levels of these two separate self-consistent calculations. Finally, $\Phi$ can be evaluated as $\Phi = {V}_{vac} - {E}_{F}^{bulk} + ({V}^{bulk} - {V}^{slab})$. This is how, we can get rid of the quantum-size effect in designing slabs for work-function calculation \cite{wf_theo}. Furthermore, for the semi-infinite systems, the band gap is not much dependent on the surface, and for intrinsic semiconductors the $E_F^{bulk}$ lies at the mid-point of VBM and CBM; therefore, the exact position of energy bands with respect to vacuum energy can be identified. Each slab is created with four layers and inter-slab vacuum is of 15 \AA. We have provided validation test of our approach on 2D-graphene,\cite{ziegler2011} and bulk-Si \cite{hollinger1983}.}

{\textbf {Phonons for dynamical stability:~}}To check the dynamical stability in terms of phonons for $\gamma$-\cn, we use first-principles density-functional-perturbation theory (DFPT). The DFPT is a straightforward approach that permits phonons to be calculated by perturbing atomic positions on $2\times2\times2$ supercell (112 atom/cell) of the original 14 atom/cell. All the atomic coordinates are  relaxed up to 10$^{-6}$eV/\AA. The atoms are displaced by 0.01 \AA~from their equilibrium positions to calculate force constants. We use these force constants to calculate phonon dispersion along the high-symmetry direction of the Brillouin zone of $\gamma$-phase \cite{phonopy}.

\section{Results and Discussion}
We discuss band-structure of 2D graphene \cite{Cabon_1}, and bulk-Si \cite{Si_expt} calculated using FP-NMTO-LDA+vLB to showcase the applicability of our approach and compared with other theory and experiments. We then study allotropes of \cn, both non-graphitic and graphitic phases, for structural and electronic properties. 

%%% FIGURE 1
\begin{figure}[t]
\centering
\includegraphics[scale=0.265]{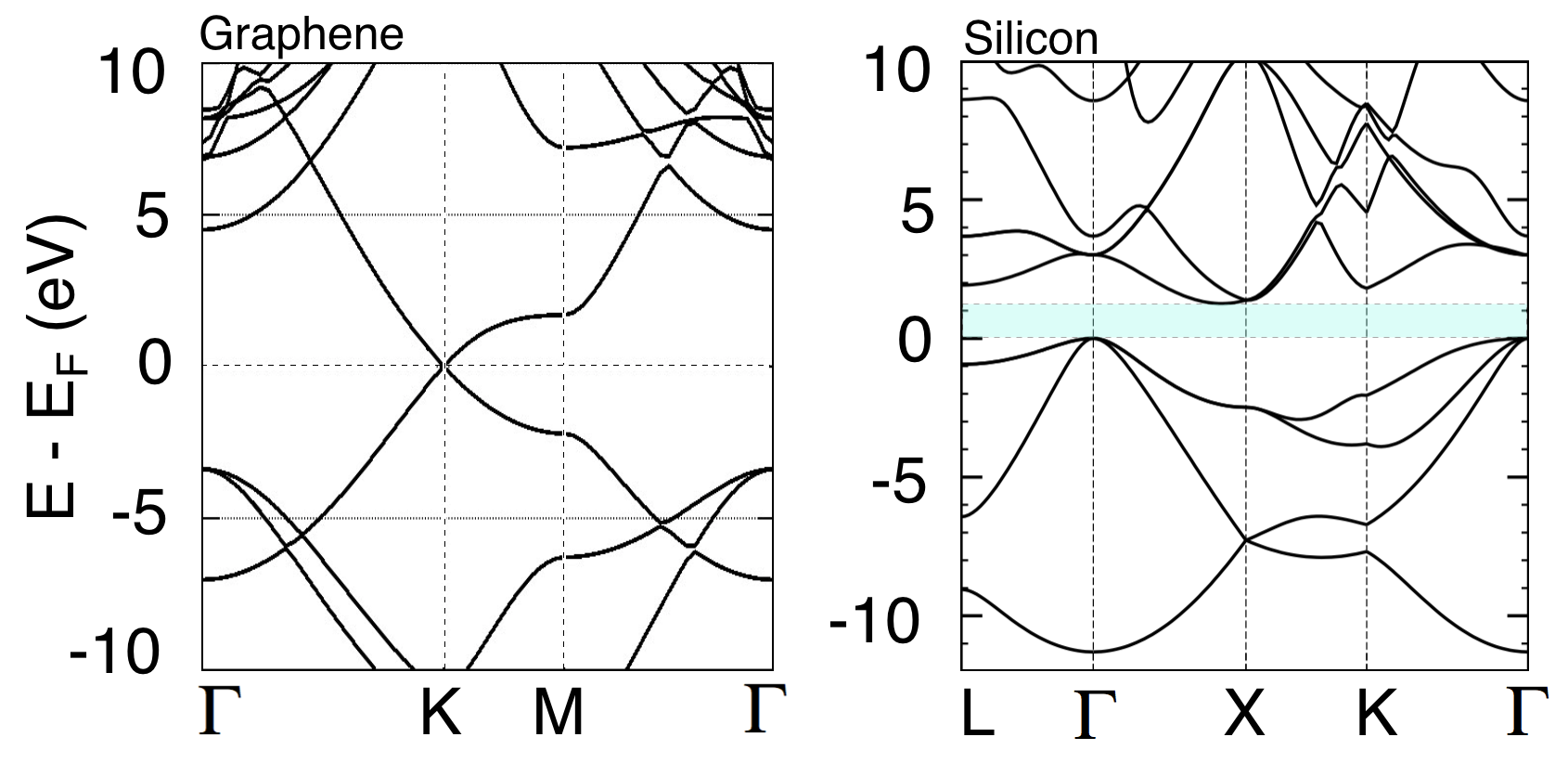}
\caption{FP-NMTO-LDA+vLB calculated band-structure of 2D-graphene (left)  and bulk-Si (right). The predicted Dirac-point in 2D-graphene (K-point) \cite{Cabon_1} and band gap of bulk-Si (1.25 eV) is in good agreement with experiment (1.17 eV)  \cite{Si_expt} and other theory \cite{PS2013}.}
\label{fig1_new}
\end{figure}

{\textbf {Graphene:}}~With a 2D hexagonal structure, graphene is very well-studied form of carbon and considered as the building block of many carbon allotropes \cite{jana2019,Cabon_9}. The calculated band-structure  (Fig.~\ref{fig1_new}, left-panel) shows zero band gap at the K-point with Dirac (Fermionic) linear dispersion. Here, the zero band gap is attributed to the sub-lattice symmetry of the 2D graphene.

{\textbf {Bulk Si}:~}~The predicted band gap of bulk-Si (1.25 eV) using LDA+vLB (Fig.~\ref{fig1_new}, right-panel) is also in good agreement with hybrid-functional (1.24 eV) and experiments (1.17 eV) \cite{Si_expt}. Notably, we reproduce the indirect ($\Gamma-X$) nature of Si band gap  \cite{PS2013}, which is not found by standard semi-local XC, for example LDA/GGA \cite{PBE,PBEsol}.

\subsection{\cn~polymorphs}
\begin{figure}[t]
\centering
\includegraphics[scale=0.33]{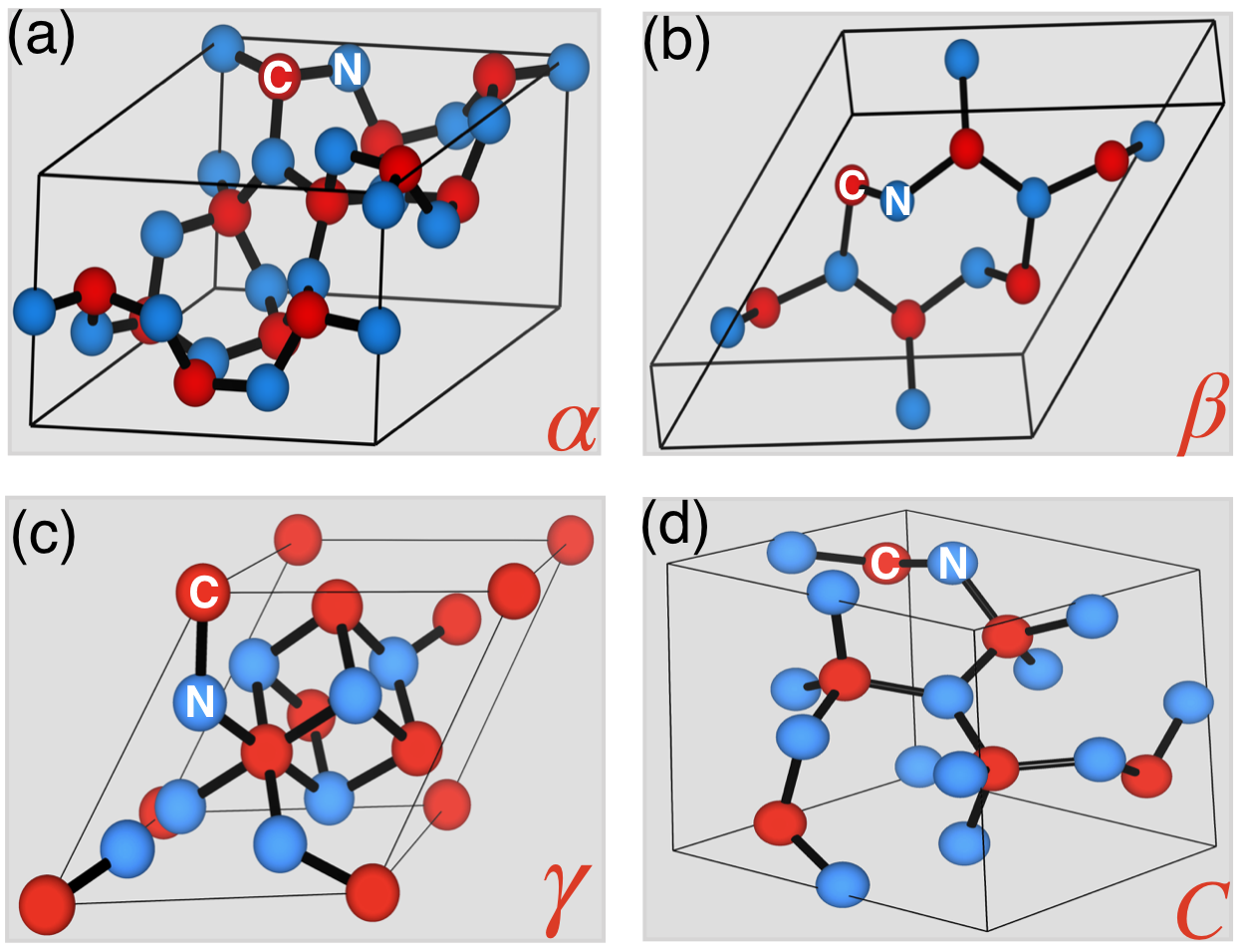}
\caption{Crystal structure of carbon-nitride (\cn) polymorphs: (a) $\alpha$-phase (trigonal; space group:$P3_1c$;159); (b) $\beta$-phase (hexagonal; $P6_3/m$; 176); (c) $\gamma$-phase (spinel;  $Fd\bar{3}m$; 227); (d) C-phase (cubic; $I43d$; 220). See phase stability in Fig. S3.}
\label{fig1}
\end{figure}

{\textbf {Non-Graphitic Phases -- structural property}:~} An account of $\alpha, \beta$, $\gamma$ (spinel), and cubic phases of \cn ~has been presented earlier \cite{Hemley1996,Molina1999}. The structures are shown in Fig.~\ref{fig1}. 
The  $\alpha$-\cn~phase (Fig.~\ref{fig1}(a)) consists of the corner sharing CN$_4$ tetrahedron along with pyramidal NC$_3$ in spheroidal cavities, suggesting that C and N are $sp^3$ and $sp^2$ hybridized, respectively \cite{Hemley1996}. The $\alpha$-phase has 4 formula units (f.u.) per cell with 28 atoms. 
The $\beta$-\cn ~(Fig.~\ref{fig1}(b)) has 6-, 8- and 12-membered rings of alternating C and N atoms \cite{Molina1999}. The $\beta$-phase has two inequivalent N atoms: (I) one N has three equidistant nearest carbon atoms with the C-N-C angles $\sim$120$^\circ$, which shows that the C and N are $sp^2$ hybridized; and (II)  other N is bonded to three non-planar carbon atoms in $sp^{2}$-$sp^{3}$ fashion.
The $\gamma$-phase {(Fig.~\ref{fig1}(c))} has a spinel structure, which is known as cubic-modification of boron-nitride (c-BN) \cite{Mo1999}. Two carbon atoms bond octahedrally to six nitrogen atoms,  while the third carbon atom bonds tetrahedrally to four nitrogen atoms \cite{Jiang2001}. The tetrahedrally and the octahedrally  bonded C-N are arranged alternatively in a 1$:$2 ratio. These are sequentially connected to one another at the N$-$corners. 
The cubic-\cn~{(Fig.~\ref{fig1}(d))} is a high-pressure modification of Willemite mineral (Zn$_2$SiO$_4$) \cite{Hemley1996}, formed by replacing the O with N and Zn and Si with C of Zn$_2$SiO$_4$ structure, so N and C are $sp^{3}$ hybridized. 

%%% FIGURE 2

\begin{table*}[t]
	\begin{adjustbox}{width=\textwidth,center}
		\begin{tabular}{c|ccc|cccc|ccc|ccc}
			\hline  
			& \multicolumn{3}{c|}{\large\textbf{$\alpha$}}                                                      & \multicolumn{4}{c|}{\large\textbf{$\beta$}}                                                                                               & \multicolumn{3}{c|}{\large\textbf{$\gamma$}}                                                      & \multicolumn{3}{c}{\textbf{Cubic}}                                                         \\  
			\multirow{-2}{*}{\textbf{Polymorphs}}                    & \multicolumn{3}{c|}{\bf{$P3_1c$ (159)}}                                                          & \multicolumn{4}{c|}{\bf $P6_3/m$ (176)}                                                                                                 & \multicolumn{3}{c|}{\bf $Fd\bar{3}m$ (227)}                                                     & \multicolumn{3}{c}{\bf $I\bar{4}3d$ (220)}                                                     \\ \hline
			\textbf{XC$\rightarrow$}                                                 & \cellcolor[HTML]{EFEFEF}{\color[HTML]{333333} \textbf{vLB}} & \textbf{PBE} & \textbf{PBEsol} & \cellcolor[HTML]{EFEFEF}{\color[HTML]{333333} \textbf{vLB}} & \textbf{PBE} & \textbf{PBEsol} & {\color[HTML]{333333} \textbf{Expt.}} & \cellcolor[HTML]{EFEFEF}{\color[HTML]{333333} \textbf{vLB}} & \textbf{PBE} & \textbf{PBEsol} & \cellcolor[HTML]{EFEFEF}{\color[HTML]{333333} \textbf{vLB}} & \textbf{PBE} & \textbf{PBEsol} \\ \hline
			a (\AA)                                                                                     & \cellcolor[HTML]{EFEFEF}{\color[HTML]{333333} 6.4000}      & 6.5350       & 6.4945          & \cellcolor[HTML]{EFEFEF}{\color[HTML]{333333} 6.4100}      & 6.4538       & 6.4177          & 6.4017                                & \cellcolor[HTML]{EFEFEF}{\color[HTML]{333333} 6.7824}      & 6.7867       & 6.7350          & \cellcolor[HTML]{EFEFEF}{\color[HTML]{333333} 5.4410}      & 5.4496       & 5.4169          \\ \hline
			c (\AA)                                                                                     & \cellcolor[HTML]{EFEFEF}{\color[HTML]{333333} 4.6296}      & 4.7382       & 4.7099          & \cellcolor[HTML]{EFEFEF}{\color[HTML]{333333} 2.4537}      & 2.4243       & 2.4103          & 2.4041                                 & \cellcolor[HTML]{EFEFEF}{\color[HTML]{333333} -}           & -            & -               & \cellcolor[HTML]{EFEFEF}{\color[HTML]{333333} -}           & -            & -               \\ \hline
			a/c                                                                                     & \cellcolor[HTML]{EFEFEF}{\color[HTML]{333333} 1.3824}      & 1.3792       & 1.3789          & \cellcolor[HTML]{EFEFEF}{\color[HTML]{333333} 2.6124}      & 2.6621       & 2.6626          & 2.6628                                 & \cellcolor[HTML]{EFEFEF}{\color[HTML]{333333} -}           & -            & -               & \cellcolor[HTML]{EFEFEF}{\color[HTML]{333333} -}           & -            & -               \\ \hline
			V (\AA$^3$)                                                                                 & \cellcolor[HTML]{EFEFEF}{\color[HTML]{333333} 165.766}     & 175.242      & 172.041         & \cellcolor[HTML]{EFEFEF}{\color[HTML]{333333} 87.311}      & 87.450       & 85.973          & -                                     & \cellcolor[HTML]{EFEFEF}{\color[HTML]{333333} 77.999}      & 78.149       & 76.375          & \cellcolor[HTML]{EFEFEF}{\color[HTML]{333333} 80.539}      & 80.958       & 79.510          \\  \hline
			\begin{tabular}[c]{@{}c@{}}$\rho$\\  (gm/$cm^3$)\end{tabular}                               & \cellcolor[HTML]{EFEFEF}{\color[HTML]{333333} 3.6887}      & 3.4893       & 3.5542          & \cellcolor[HTML]{EFEFEF}{\color[HTML]{333333} 3.6114}      & 3.4961       & 3.5562          & -                                     & \cellcolor[HTML]{EFEFEF}{\color[HTML]{333333} 3.9197}      & 3.9122       & 4.0031          & \cellcolor[HTML]{EFEFEF}{\color[HTML]{333333} 3.7962}      & 3.7765       & 3.8452          \\ \hline
			$B_0$ (GPa)                                                                                   & \cellcolor[HTML]{EFEFEF}{\color[HTML]{333333} 452.676}     & 419.362      & 453.841         & \cellcolor[HTML]{EFEFEF}{\color[HTML]{333333} 590.139}     & 413.894      & 440.825         & -                                     & \cellcolor[HTML]{EFEFEF}{\color[HTML]{333333} 421.482}     & 391.717      & 421.015         & \cellcolor[HTML]{EFEFEF}{\color[HTML]{333333} 469.360}     & 442.050      & 467.965         \\ \hline
		\end{tabular}
	\end{adjustbox}
	\caption{Comparison of lattice constants and bulk moduli of \cn~ polymorphs calculated using vLB, PBE and PBEsol functionals. Experimental lattice constant of $\beta$ phase \cite{doi:10.1021/acsami.6b11427} are compared with calculated values. Bulk moduli calculated with vLB and PBEsol are in good agreement.}
	\label{tab1}
\end{table*}

{\par} The structural properties of \cn ~polymorphs are summarized in Table.~\ref{tab1}.~The PBE, PBEsol and LDA+vLB calculated structural parameters are in good agreement \cite{Yang2007, doi:10.1021/acsami.6b11427}. Our predicted lattice constants are similar to those of experiment for $\alpha$ ($a$ = 6.5680 \AA, $c$ = 4.7060  \AA) and $\beta$ phases (a=6.4017 \AA, c=2.4041 \AA) \cite{qiu2003, doi:10.1021/acsami.6b11427}. We also evaluate bulk moduli by fitting lattice constant versus total energies to the Birch-Murnaghan equation of state \cite{MurnBM,BirchBM}, which indicates that \cn ~undergoes uniform compression under applied hydrostatic pressure. The agreement between LDA+vLB and PBEsol calculated numbers establishes effectiveness and the utility of vLB-modified XC functional \cite{PS2016}.  The electronic and optical property calculations are carried out using optimal lattice parameters given in Table.~\ref{tab1}. 

\begin{figure}[t]
\centering
\includegraphics[scale=0.265]{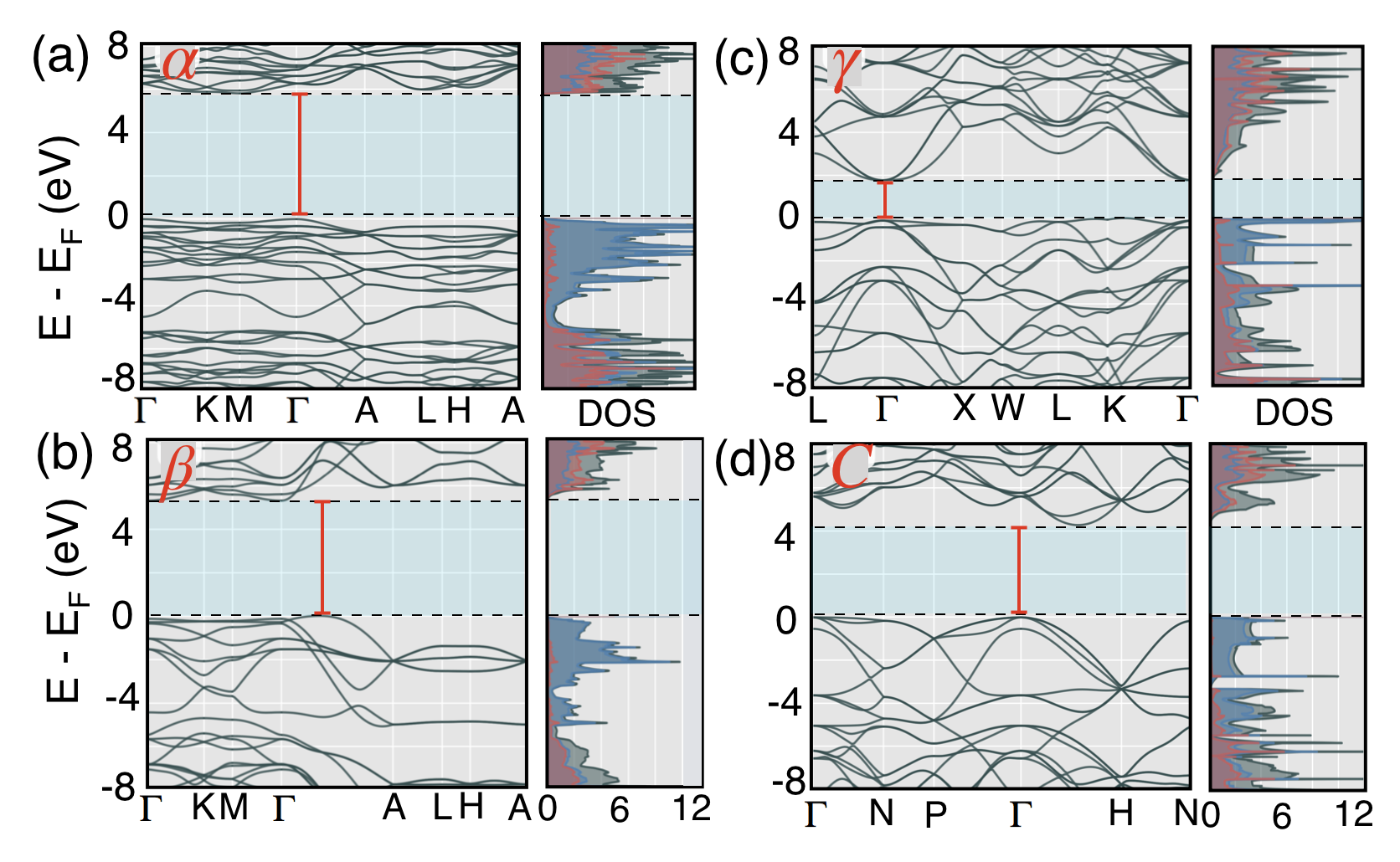}
\caption{Electronic dispersion and DOS of \cn~polymorphs: (a) $\alpha$; (b)  $\beta$; (c) $\gamma$, and (d) $C$ phases. All phases show indirect gaps, except for $\gamma$. C (red line) and N (blue line) projected DOS are shown.}
\label{fig2}
\end{figure}

{\textbf {Non-Graphitic Phases -- electronic structure}:~}Band structure is one of the most stringent tests to detail the materials physics. For example, Si, calcite and Cu have similar electron densities, but they have very different physical and electronic properties. This drives us to understand the band structure versus energy of \cn ~polymorphs in detail, shown in Fig~\ref{fig2} with a zero of energy at the VBM (E$_{F}$). The LDA+vLB band gaps for $\alpha$, $\beta$, $\gamma$ and cubic ($C$) phases are $5.81$, $5.32$, $1.81$ and $4.23$~eV, respectively. The band gaps of $\alpha$, $\beta$ and $C$ phases are indirect in nature, while $\gamma$ phase shows a direct gap.  We find good agreement between our predictions and more advanced DFT techniques (e.g., HSE and $G_0W_0$), see Table.~\ref{tab2}.

{\par} The projected and total density of states (DOS) of N and C are shown in Fig. \ref{fig2} for each of the \cn ~polymorphs. In Fig.~\ref{fig2}~(a), for the $\alpha$ phase the valence bands (VBs) below $-4$ eV phase are dominated by both N and C, while, the bands from $-4$ eV to E$_{F}$ are mostly from N. This suggests that the electrons of N are loosely bound than those of C. The $\alpha$ phase has a wider band gap, and steeper VB and CB edges, compared to the others. The VB maxima and CB minima in $\alpha$ phase are at $\Gamma$ and M point, respectively. Similar to $\alpha$, we find that $\beta$ ($\Gamma-A$ to $\Gamma$) and $C$ ($\Gamma$ to $\Gamma-H$) are indirect band gap semiconductor with no contribution from C bands near E$_{F}$. However, the $\gamma$ phase is a direct ($\Gamma$-$\Gamma$) band gap with minor contribution from C bands near E$_{F}$. The electronic structures of $\alpha$, $\beta$, $\gamma$ and cubic polymorphs split to form the separate VB, which reflects the presence of smaller asymmetric part of the potential. The presence of asymmetric potential is the reason behind the stronger mixing of low lying N and C states.

\begin {table*}[t]
	\begin{adjustbox}{width=\textwidth,center}
\begin {tabular}{c|c|cccc|ccccc}\hline 
{}& \multicolumn{10}{c}{\large \bf Band Gap (eV)}\\ \hline

{\bf \cn } & {\bf Expt.}&\multicolumn{4}{c}{\bf This Work} & \multicolumn{5}{|c}{\bf Others}\\ 
 Polymorphs		&  	& LDA       & PBE &  vLB & HSE 	& LDA \cite{Molina1999,crok,yao,Mattesini2000,Xu2012} & PBE & HSE\cite{Makaremi2018} & $GW$\cite{Xu2012} & mBJ\cite{Reshak2014}  \\ \hline     
$\alpha$    &  -- &5.24$^*$&--& 5.81 & 5.81	& 3.8-3.85   	&   --    &  --   &     5.49 & --\\
$\beta$     &   --	&3.27       &3.31& 5.32 & 5.34	& 3.11-3.56    	&   --    & --  & 4.85, 6.4\cite{crok} & --\\
$\gamma$ &  --	&1.19       &1.14& 1.81 & 1.95	&   -- 	&  --   	&  --  &  2.01 {\bf(this work)} & -- \\
$cubic$     &   --	&2.86       &2.95& 4.23 & 4.43	& 2.90-2.91  	&   --  	&   --   &  4.30 & --\\
$t-g(AA)$  & 3.1\cite{Khab2000}&2.79&	2.70 & 3.02	& 3.21 &  1.16-1.48,0.699\cite{Reshak2014}  &  0.870\cite{Reshak2014}   &  3.19  & 2.97 & 2.549\\
$t-g(AB)$  & -- &2.66 & 2.67& 2.95 & 3.39 &  1.204\cite{Reshak2014}  & 1.357\cite{Reshak2014} & --   &--  & 2.99\\
$h-g$     & 2.67-2.95 \cite{Wang2009,Wang2009_1,Wang2010,Mattesini2000,gcn} & --& --	&  2.71 & 2.71 &  --  & -- & 2.772   & 2.88 & --\\
\hline 
\end {tabular}
\end{adjustbox}
\caption{Calculated band gaps for \cn ~polymorphs using (LDA, PBE, vLB, HSE) and comparison with other theories \cite{Molina1999,crok,yao,Mattesini2000,Xu2012,Makaremi2018,Reshak2014} (LDA, PBE, HSE, GW, mBJ) and experiments \cite{Khab2000,Wang2009,Wang2009_1,Mattesini2000,gcn}. HSE calculations are done with plane-wave basis.\cite{QE1,QE2} Band gaps using vLB \cite{PS2016} and HSE (this work) shows good agreement with existing HSE, GW and experiments. We also include non-self-consistent GW band gap for $\gamma$-\cn ~phase calculated with QE package. Band gaps from HSE (1.95 eV), GW (2.01 eV) and vLB$+$LDA (1.81 eV) are in good agreement.}
\label{tab2}
\end {table*}

%%% FIGURE 3

\begin{figure}[t]
\centering
\includegraphics[scale=0.29]{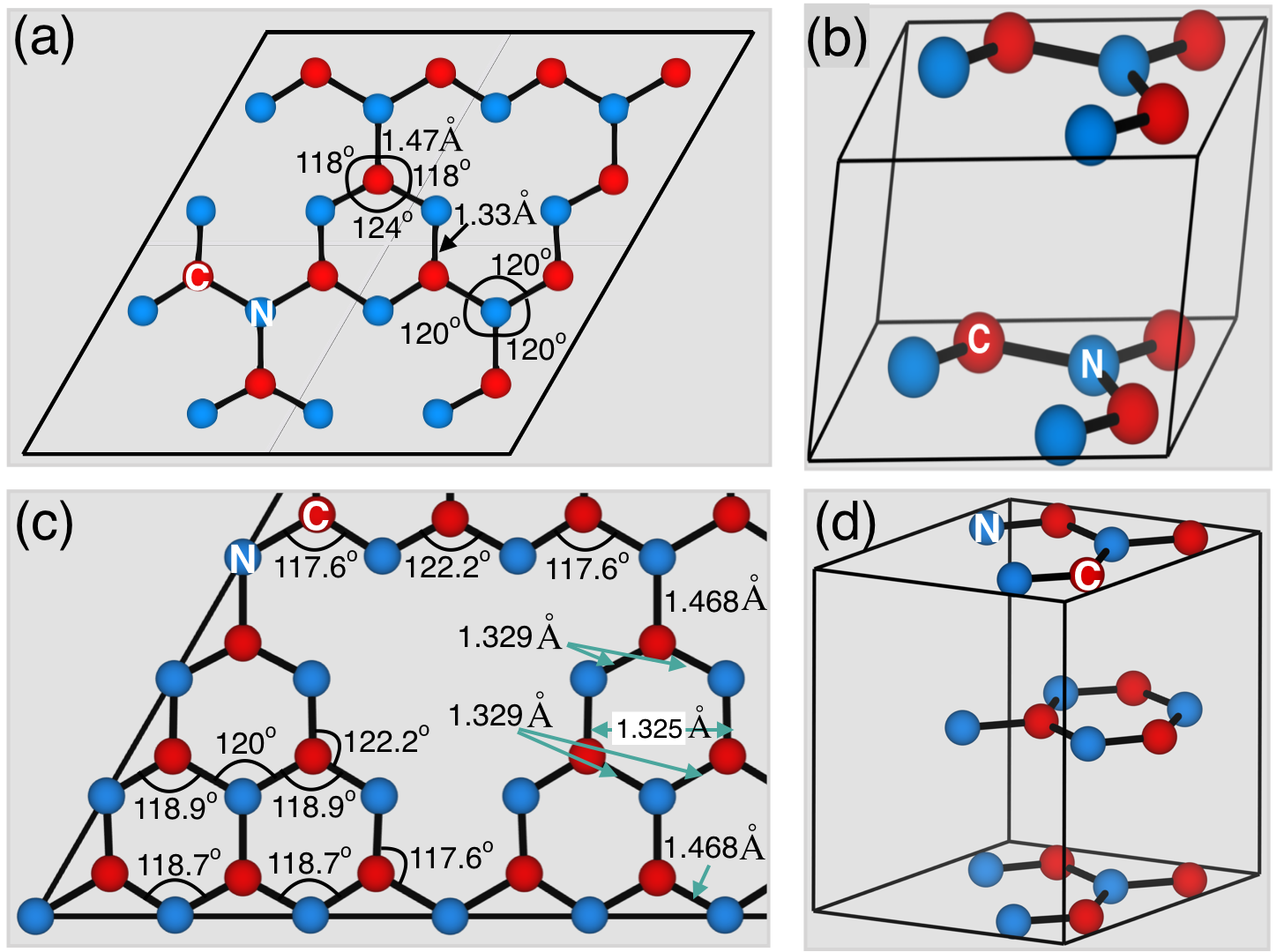}
\caption{Crystal structure of relaxed graphitic-\cn: (a) $t-g$(AA), {(b) $t-g$(AA) unit cell, and (c) $h-g$(AA), (d) $t-g$(AB) unit cell} \cn. For relaxed $t-g$ phase, bond angles and lengths vary from ($118-124^o$) and ($1.33-1.47$ \AA), respectively, while, in relaxed $h-g$ phase, bond angles and lengths vary as ($117.6-122.2^o$) and ($1.325-1.468$ \AA).}
\label{fig3}
\end{figure}

{\textbf {Graphitic Phases -- structural property}:~}~Notably, graphitic phases of \cn ~are considered for next generation visible-light-driven metal-free, non-toxic, earth-abundant semiconductors, which can find applications in energy conversion, hydrogen evolution, sensing and imaging \cite{Wang2009}. Earth-abundant graphitic \cn ~possesses excellent electronic band structures, electron-rich properties, and higher stability. The $sp^2$ hybridized planar graphitic \cn~(Space group: $P6m2$; \#187) in Fig.~\ref{fig3} can be viewed as graphite whose C lattices are partially substituted with nitrogen (in regular fashion) \cite{Wang2009}. We prepare triazine with (AA; AB) stacking and  heptazine with AA stacking, respectively. 
The large cavities in t-g(AA), t-g(AB) and h-g(AA) allow neighboring atoms to relax and lead to increased bond lengths and angles compared to other phases (Fig. \ref{fig3}). For t-g(AA), the equilibrium lattice constants are $(a, c)=4.786, 3.758$ \AA, where  bond lengths and angles vary from ($1.327-1.463$ \AA) and ($118-124^{o}$), respectively. The calculated inter-layer spacing of $3.505$ \AA~and lattice constant of t-g(AB) $(a,c)=(4.784, 7.007)$ \AA ~is in good agreement with experiment ($(a,c)=(5.041, 6.576)$ \AA) \cite{algara2014}. The bond lengths vary from $1.326-1.468$ \AA, while the angles range from $117.6-122.2^{o}$. Due to larger cavity size compared to t-g(AA) and t-g(AB), the h$-$g (AA) structure shows larger degree of relaxation, both in bond lengths and angles as shown in Fig.~\ref{fig3}(d).

%%% FIGURE 4

{\textbf {Graphitic Phases -- electronic structure}:~} Experimentally thin films of graphitic-\cn ~can be produced on different substrates. By controlling the synthesis conditions, the band gap can be tuned from~$2.65$ to $3.1$ eV, which falls in the range of visible light \cite{Khab2000,Wang2009,Mattesini2000}. The LDA+vLB band gap of 3.02 eV for AA (trigine) and 2.95 eV for AB (heptazine) stackings are in close agreement with experimental gap ($3.1$ eV) \cite{Khab2000}. The QE-HSE predicted band gap for AA  and AB stackings is $3.21$ eV and $3.39$ eV, respectively. LDA+vLB predicted band gaps are in excellent agreement with HSE, other theory \cite{Makaremi2018,Xu2012}, and other experiments. The t-g(AA) shows indirect (K$-$A) band gap, whereas t-g(AB) is direct. The LDA+vLB  predicted indirect ($\Gamma$-M) band gap ($2.71$ eV) of h-g-\cn~is in good agreement with HSE ($2.71$ eV), other theory \cite{Xu2012,Makaremi2018}, and experiments $2.67-2.95$ eV \cite{Wang2009,Wang2009_1,Wang2010,Mattesini2000,gcn}.

%%% FIGURE 5
\begin{figure}[t]
\centering
\includegraphics[scale=0.28]{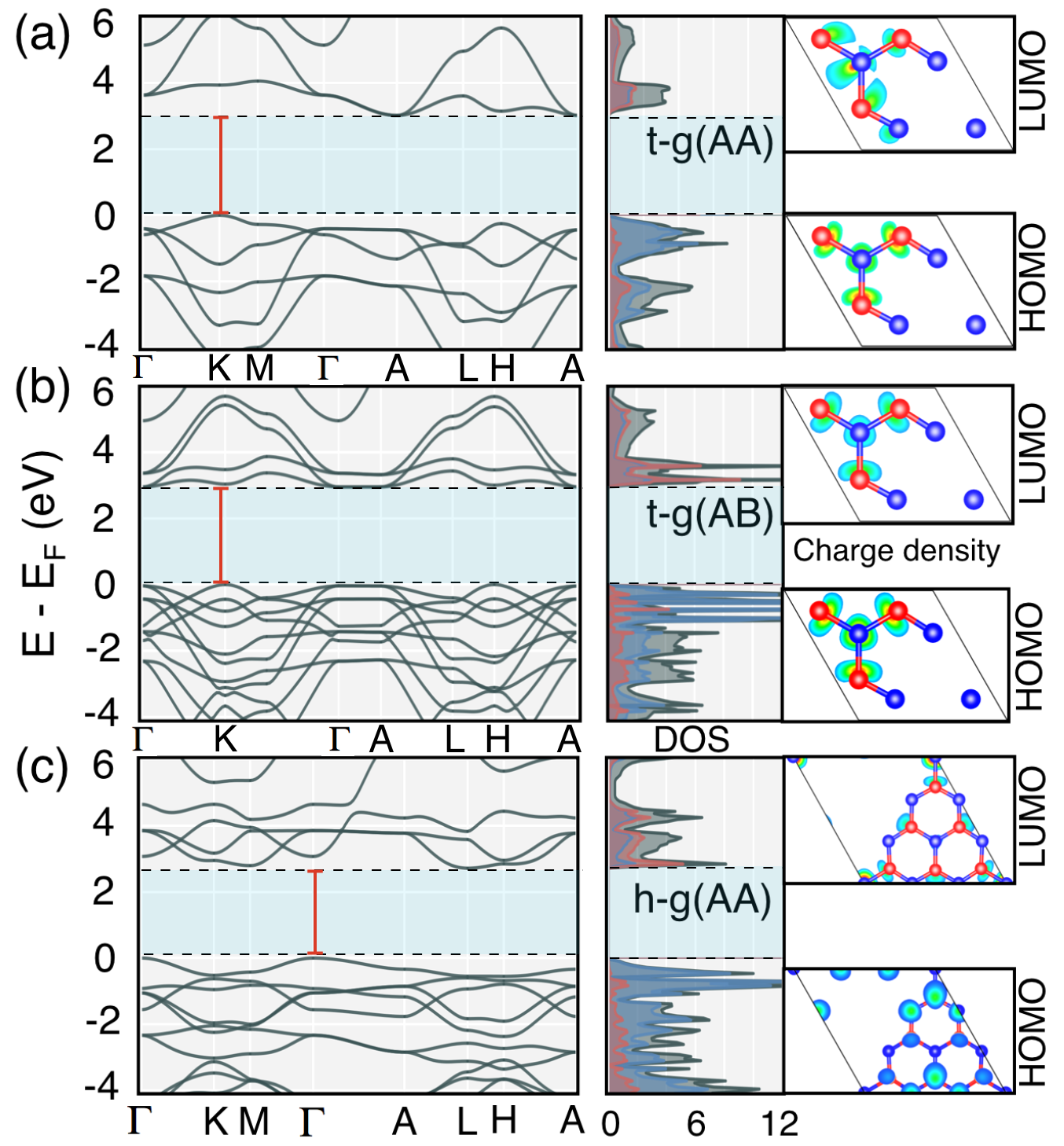}\\ 
\caption{Electronic band-structures, density of states and charge densities (VBM \& CBM states) of relaxed graphitic-\cn -- (a) $t-g$(AA), (b) $t-g$(AB), and (c) $h-g$(AA). The $t-g$(AA), $t-g$(AB) and $h-g$(AA) phases exhibit band gaps of $3.02$ eV (K--A; indirect), $2.95$ eV ($\Gamma-\Gamma$; direct) and $2.71$ eV ($\Gamma$-M; indirect), respectively. C-$p$ (red) and N-$p$ (blue) projected density of states (PDOS) are shown. Charge-density isosurface plots are set at  $0.035$ electrons/\AA$^{3}$.}
\label{fig4}
\end{figure}

{\par}To identify active sites in \cn ~polymorphs, we plot the CBM and VBM charge densities (Fig.~\ref{fig4}, right-panel). The CBM of $t-g$(AA), $t-g$(AB) and $h-g$(AA) \cn~are mainly composed of N-$2p$ and C-$2p$ states, while the VBM is dominated by N-$2p$ states. Also, strongly localized CBM and VBM states in charge density plots suggest the possibility of low photo-absorption efficiencies, i.e., $e^{-}-h^{+}$ excitation under visible-light irradiation occuring on the edge of N and C atoms. 

{\par}In Table~\ref{tab2}, we compare the calculated band gaps from LDA, PBE, vLB, and HSE (present work) with other theory and experiments \cite{Molina1999,crok,yao,Mattesini2000,Xu2012,Makaremi2018,Reshak2014,Khab2000,Wang2009,Wang2010,Mattesini2000,gcn}. The LDA+vLB and QE-HSE predicted band gap for cubic phase is in good agreement with GW, while band gap for $\alpha$ and $\beta$ phase is higher than GW. 

\subsection{Effective mass -- \cn~polymorphs}~
\begin{table}[]
	\centering
	\begin{tabular}{|c|c|c|}\hline
		{\bf Polymorphs} & \multicolumn{2}{c|}{\bf m$^{\ast}$} \\ \hline
		& {\bf e$^{-}$}            & {\bf h$^{+}$}           \\ \hline
		$\alpha$   &      	0.045              &    -0.050             \\
		$\beta$    &     	0.022               & 	-0.137			\\
		$\gamma$   &       	0.016              &      -0.065          \\
		$cubic$    &    	0.039                & 	-0.065			\\
		$t-g(AA)$   &      	0.044              &     -0.044           \\
		$t-g(AB)$    &      0.028              &    	 -0.026	\\
		$h-g(AA)$   &       0.045               &      -0.081       \\ \hline              
	\end{tabular}
	\caption{Effective mass (m$^{\ast}$) of \cn ~polymorphs in  electron rest mass ($m_0$) units.}
	\label{tab_m}
\end{table}
The dynamical activity of the charge careers in semiconductor depends on its mobility, which is inversely proportional to the effective mass (m$^{\ast}$). Near the band edges, effective mass is described by $m^{\ast}={\hbar^2} / \left( \frac{d^{2}E}{dk^{2}} \right)$, where $\frac{d^{2}E}{dk^{2}}$ is the band-curvature at VBM for holes and at CBM for electrons. The m$^{\ast}$ for \cn~polymorphs is shown in {Table~\ref{tab_m}}. The calculated m$^{\ast}$ in graphitic phase is in good agreement with the available results \cite{dong2017, li2014} and we find electron m$^{\ast}$ of $\gamma$ phase is much lower than graphitic phases. This indicates that the electron mobility in $\gamma$ phase is easier. Interestingly, between the two types of stacking in t-graphitic phases, the A-B alternate stacking provides lower electron and hole effective mass and higher mobility than the AA phase. We attribute this to the lower symmetry of AB phase.

\subsection{Optical Properties:} Solar energy output is mostly dominated by: (I) ultraviolet ($\sim$5\%), visible ($\sim$45\%), and infrared ($\sim$50\%) region of electromagnetic wave spectrum \cite{36_new}. The visible-light photocatalysis therefore offers the best opportunity to utilize maximum solar energy \cite{nath2014ab}. Except for $\gamma$ and $h-g$, the \cn~polymorphs possess relatively wide band gaps (see Table.~\ref{tab2}). The optimal gap of $\gamma$ and $h-g$~\cn~ \cite{obg1} makes them promising polymeric semiconductors suited for visible-light absorption.

%%% FIGURE 6
\begin{figure}[t]
\centering
\includegraphics[scale=0.4]{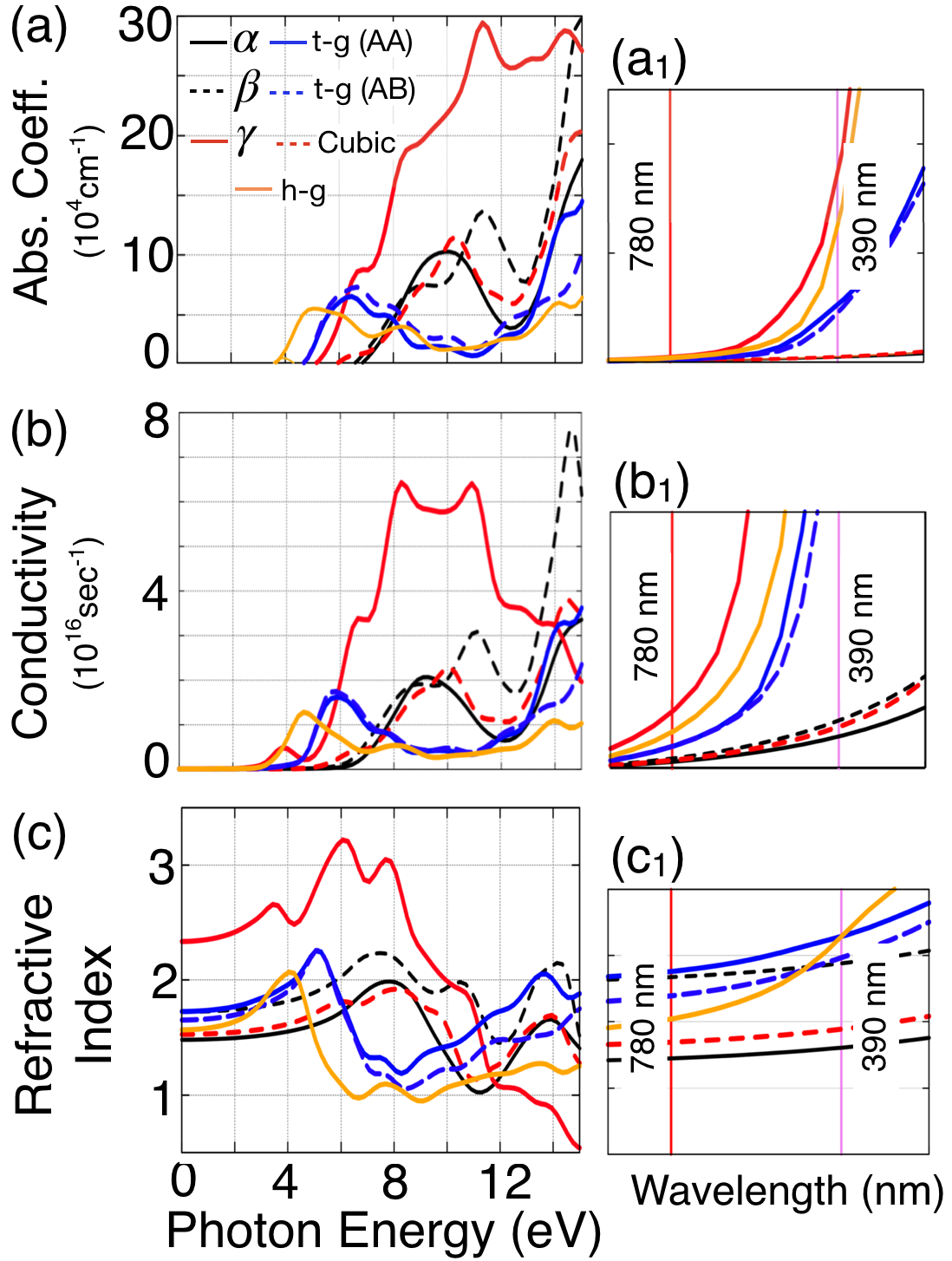} 
\caption{{QE-HSE calculated (a) Absorption spectra, (b)  optical conductivity, and (c) refractive index vs. photon energy for \cn ~polymorphs. (a$_{1}$) Absorption spectra, (b$_{1}$) optical conductivity, and (c$_{1}$) refractive index are also shown in the visible region of light.}}
\label{fig5}
\end{figure}

{\par}In Fig.~\ref{fig5}, we plot absorption coefficient, optical conductivity, and refractive index vs photon energy. Plots of effective number of electrons participating in interband transition ($n_{eff}$) shows that the (0$-$15 eV) range of photon energy is justified (Fig. S4). However, our region of interest is visible light from 1.59 eV (780~nm) to 3.18 eV (390~nm), in Fig.~\ref{fig5} (a$_{1}$, b$_{1}$, c$_{1}$). The optical spectra (obtained from the imaginary part of the dielectric function) contains the information of character and number of occupied and unoccupied bands. These together decide the accuracy of optical properties at higher excitation energies.

{\par}The absorption coefficient of a material determines the spatial region in which most of the light is absorbed. The absorption edge extent of the $\gamma$ and $h-g$ phases is relatively larger than other polymorphs due to narrower gaps of 1.81 and 2.71 eV, respectively \cite{obg1}. The onset of absorption for $\gamma$ and $h-g$ phases are located at roughly  2.0 and 3.0 eV, respectively, very close to the their band gaps. {Also, the structural change induced by relaxation in Fig. \ref{fig3}(d)  for h-g are manifested by more pronounced changes in the optical absorption in Fig. \ref{fig5}(a)\&(a1) (AA) as compared to t-g (AA) and t-g (AB). The degree of relaxation and band gap decides the absorption-onset (AO) and band-edge absorption (BE), for example, for h-g phase BE/AO is at 3.0 eV smaller than t-g(AA) and t-g(AB).} Interestingly, both $\gamma$ and $h-g$ phases show an increase in optical absorption range. However, the $\gamma$ phase shows enhanced light absorption in the whole spectral region due to direct nature of its band gap. Also, direct gap materials provide better photons to electron-hole pair conversion useful for efficient electro-optical devices -- the foremost reason behind higher optical absorbance in $\gamma$ phase, as shown in Fig.~\ref{fig5}(a). 

{\par}The accurate description of the band gap and band-positions from hybrid-functionals assure the reliability of our prediction of optical properties. We compared the QE-HSE calculated optical conductivity of \cn ~polymorphs in Fig.~\ref{fig5}(b). Our results shows that the conductivity starts with a gap, indicating the semiconducting character of each polymorph. The optical conductivity is zero below a certain energy value, which is consistent with the respective band gaps in Table 2 of \cn ~polymorphs. The optical conductivity for the $\gamma$ phase in Fig.~\ref{fig5}(b) is slightly higher than $h-g$ phase, which is related to absorption coefficient (Fig.~\ref{fig5}(a)) and shows similar behavior in the range of visible light  (1.59 to 3.18 eV). Only in the energy range of 4 to 5.5 eV does other polymorphs overtake $\gamma$, whereas it shows large jump in absorption and conductivity beyond photon energy 5.5 eV. The increased optical conductivity in visible light range makes of $\gamma$ phase a promising candidate for photovoltaic application. Refractive index is shown in Fig.~\ref{fig5}(c) as a function of photon energy (wavelength: in Fig.~\ref{fig5} (a$_{1}$, b$_{1}$, c$_{1}$)). Physically the calculated optical gap corresponds to the photon energy at which the imaginary part of the refractive index, $k$, becomes non-zero. We found that the refractive index for  \cn~polymorphs vary from $1.4$ to $1.9$ for the visible range of light $390-780$~nm, showing sensitivity of the refractive indices to crystal structure. The optical absorption spectra are directly connected to the imaginary part of the refractive index, see Eq. (4). So, the peaks and valleys in refractive index are expected in the similar energy range that of  absorption spectra.

\subsection{Work Function}
\begin{figure}[t]
\centering
\includegraphics[scale=0.27]{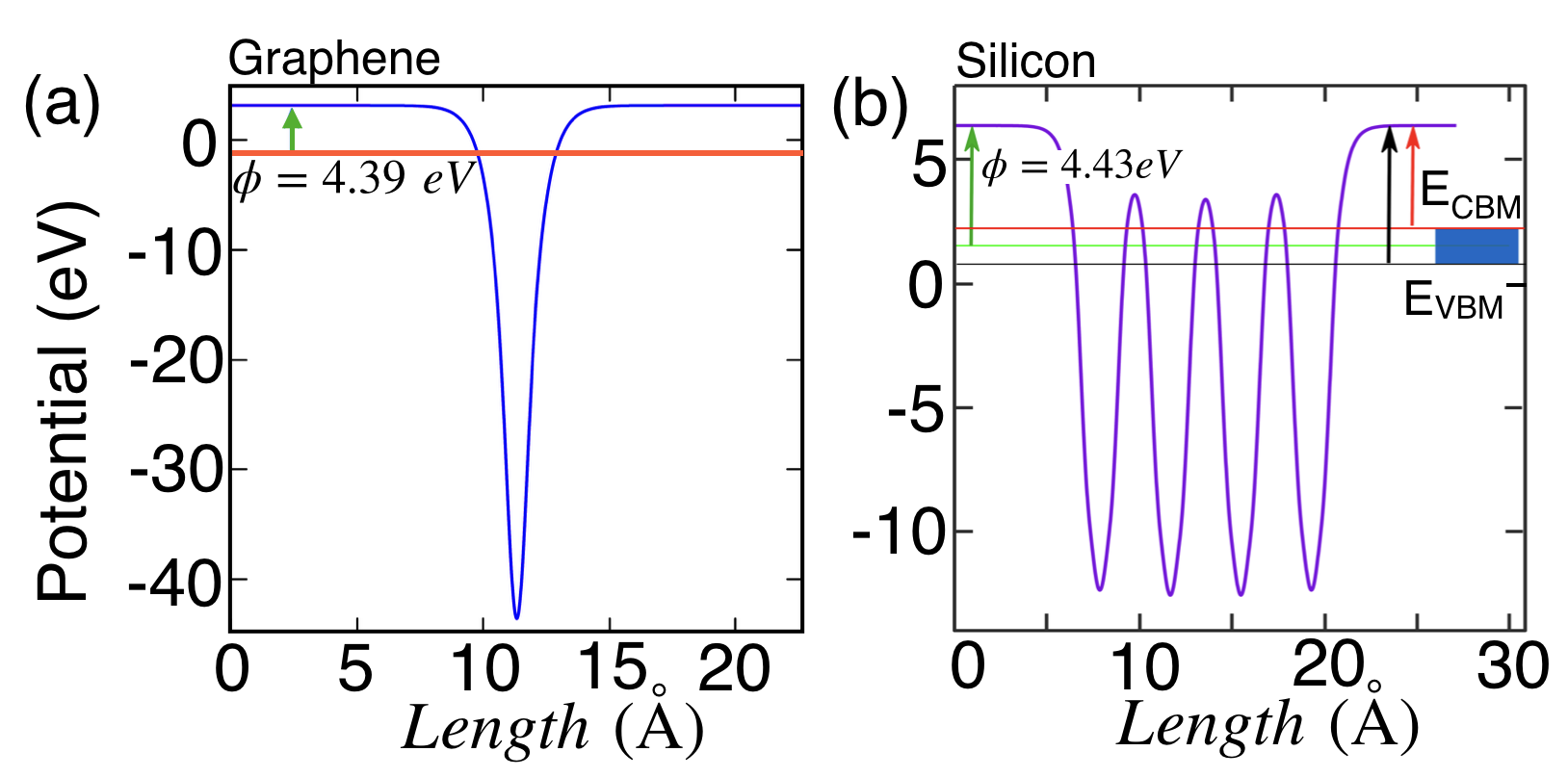}
\caption{QSE-HSE  calculated work-function of (a) 2D-graphene (4.39 eV), and (b) bulk-Si (4.43 eV) is in good agreement with experiments (2D-graphene, 4.56 eV \cite{ziegler2011}; bulk-Si, 4.87 eV \cite{hollinger1983}).}
\label{fig1_new2}
\end{figure}

{\textbf{Monolayer graphene and bulk-Si:}}~The work function, which limits the performance of devices, is an important characteristics of semiconductors that should be taken into considerations. First, we cross-validate QE-HSE method by calculating work-function of well-studied 2D-graphene, and bulk-Si. The $3\times3\times1$  and $2\times2\times1$ supercells are chosen for accurate calculations of 2D-graphene and bulk-Si, respectively. The inter-slab vacuum of 15 \AA~is chosen along (001) to avoid periodic image problem. In Fig.~\ref{fig1_new2}a, we show the calculated work-function of 4.39 eV for 2D-graphene. The $\phi_{DFT}$ is in good agreement with  $\phi_{Expt}$=4.56 eV \cite{ziegler2011}. In Fig.~\ref{fig1_new2}b, the conduction-band minima (CBM) and valence-band maxima (VBM) for Si are plotted considering E$_{vac}$= 0. The calculated work-function of 4.43 eV for Si is in very good agreement with experiments (4.87 eV) \cite{hollinger1983}. 

%%% FIGURE 7

{\textbf{\cn~polymorphs:}~}~At ambient conditions (i.e., at normal (room) temperature and one atmospheric pressure), the thermodynamic voltage for water splitting is 1.23~V \cite{meda2013,jafari2016}. But, to supply the required photovoltage, a photo-electrochemical cell with single illuminated electrode needs gaps greater than 1.6 eV \cite{bolton1985}. In Table~\ref{tab2}, we show that the gap in $\gamma$ and graphitic phases is suitable for photoabsorption. To initiate water-redox reaction in semiconductors, optimal gaps, as well as band-edge (conduction and valence) positions of VBM and CBM states are important. The valence-band maxima should be more positive ($E_{H_2O/O_2}$=1.23, 0.81 V for pH=0, 7) with respect to normal hydrogen electrode (NHE) than the water oxidation level, whereas the conduction-band minima should be more negative ($E_{H^+/H_2}$=0, $-0.41$ V for pH=0, 7 vs. NHE) than the H-production potential: 
\begin{eqnarray}
H_2O+2h^+ &\rightarrow & 2H^+ + \frac{1}{2}O_2 \nonumber\\
2H^+ + 2 e^- &\rightarrow & H_2  \quad  .
\end{eqnarray}

\begin{figure}[t]
\centering
\includegraphics[scale=0.26]{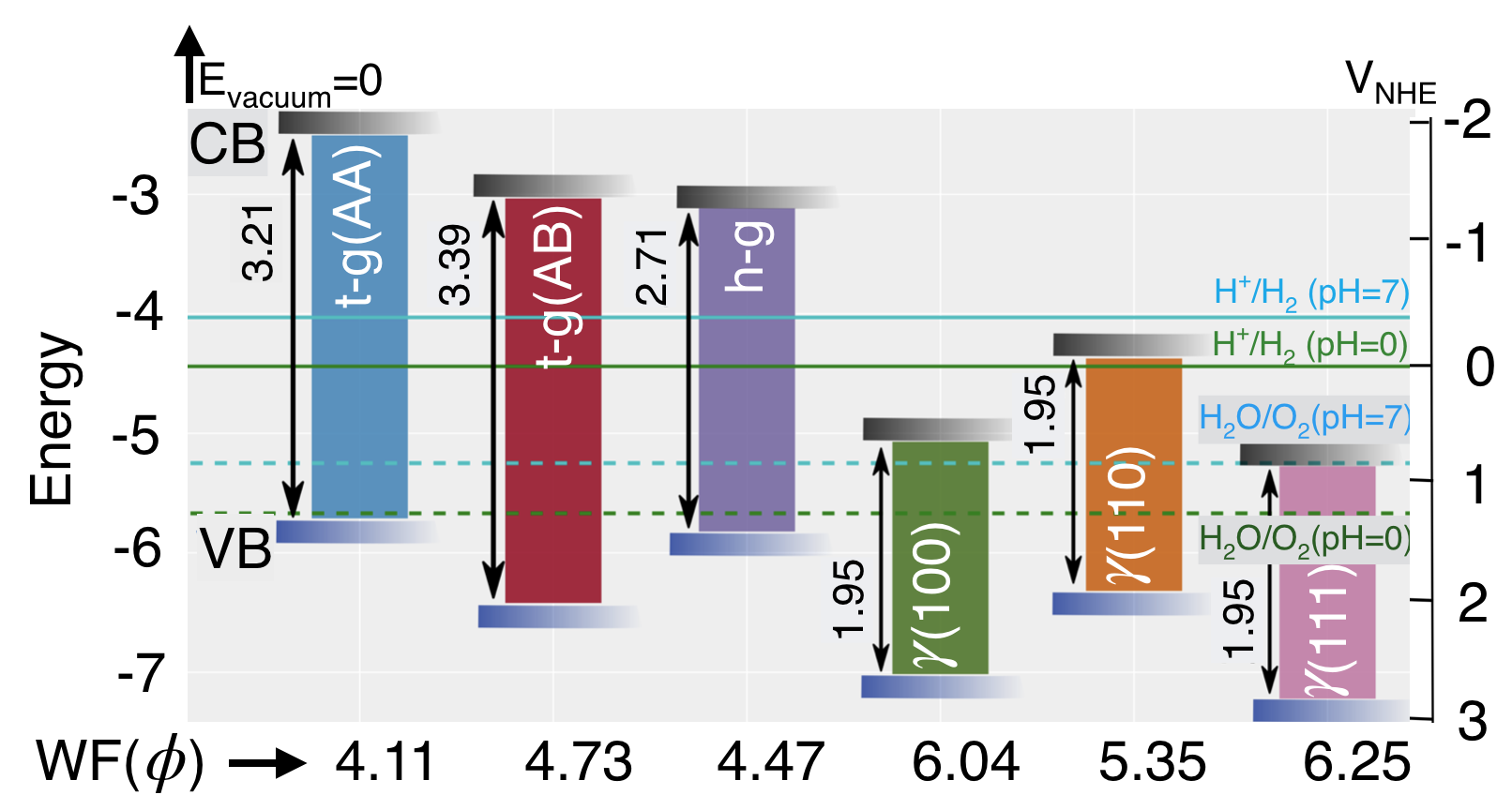} 
\caption{Conduction and valence band positions of \cn~polymorphs with respect to water reduction and oxidation potentials.}
\label{fig6}
\end{figure}

{\par}In Fig.~\ref{fig6}, we illustrate valence and conduction band positions and work function for $t-g$(AA), $t-g$(AB), $h-g$, and $\gamma$ phase for ($100$), ($110$) and ($111$) planes. The VBM for graphitic phase is positive with respect $H_{2}O/O_{2}$ -$p_{H}=0$ level while CBM are negative compared to $H^+/H_2 - p_{H}=0$ level in the entire $pH$-range (0 to 7). The work function of the triazine phases depends largely on the choice of stacking: alternate (AB) stacking increases the band gap and also enhances the work-function potential. The band-position in both t-g (AA) and t-g (AB) makes them good candidate for photocatalysis, however, t-g (AB) is more favorable due to larger work function in the entire pH range. The QE-HSE predicted $h-g$ \cn~work function (4.47 eV) agrees well  with the experiments (4.3 eV) \cite{8_2,ong}. For $h-g$-\cn, Yu \etal \cite{yu2019} showed that the band gap range of 2.44-2.69 eV~and VBM at 1.50 eV with respect to NHE at room temperature, in good agreement with our predicted band gap of 2.71 eV and VBM at 1.385 eV. The CBM and VBM positions in Fig.~\ref{fig6} suggests that graphitic-\cn ~favors water (pH = 7) over acidic water (pH $<$ 7) for photocatalysis.

%%% FIGURE 8

{\textbf{$\gamma$-\cn~workfunction:}~}Our ability to modulate the work function of semiconductor alloys, e.g., through control of surface orientation, is an enabling factor. The optimal band gap (1.95 eV) for $\gamma$-phase makes it more efficient for the photo-absorption in visible spectra compared to other \cn~polymorphs. Among the three illuminated surface conditions of $\gamma$ phase, i.e., (100), (110), (111) (see Fig. S5 in supplement), the VB and CB band range of (110) surface is {favorable at pH=0}. Our study suggest $\gamma$-\cn~as a potential candidate for photocatalytic application, a new and efficient photocatalysts from spinel group \cite{spinel_review}.

\subsection{Pressure effect on key properties of $\gamma$-C$_{3}$N$_{4}$:}
%%% FIGURE 09
\begin{figure}[t]
	\centering
	\includegraphics[scale=0.4]{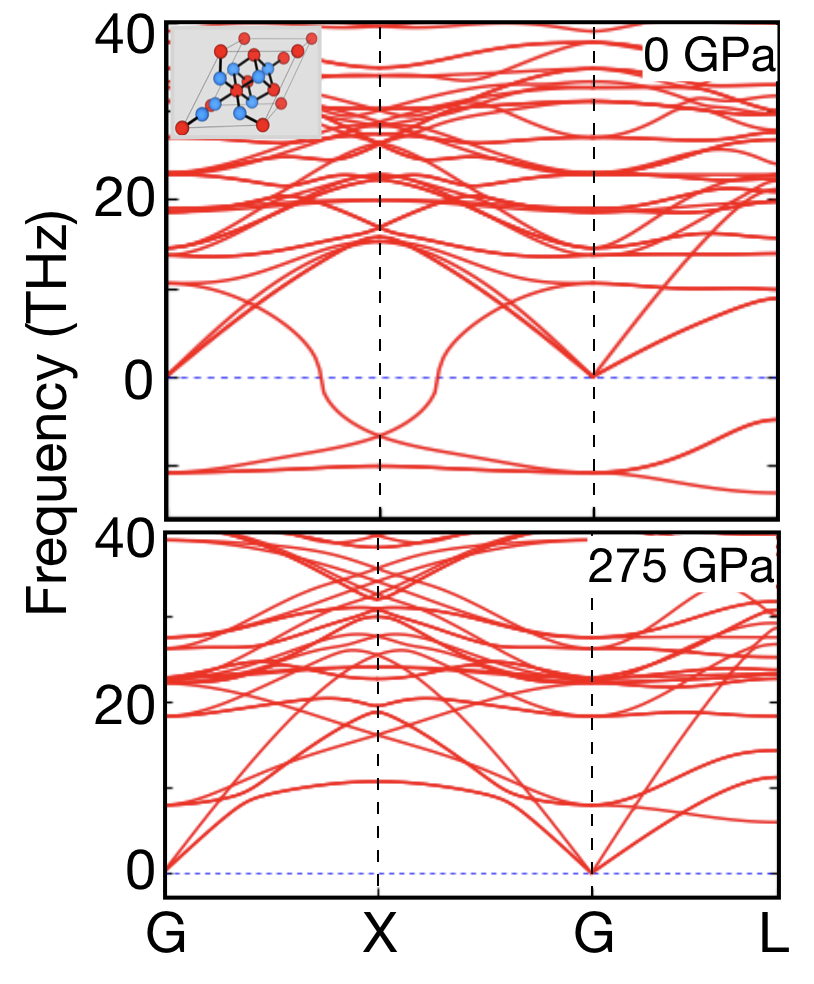} 
	\caption{Phonon dispersion in $\gamma$-\cn~along high-symmetry directions. (Top) Modes at 0~GPa shows dynamical instability, i.e., imaginary frequencies denoted by negative values. (Bottom) With pressure unstable modes disappear at/above the critical pressure of 275 GPa, with a 11.5\% volume change compared to zero pressure.}
	\label{fig7}
\end{figure}  
Optimal band gap (1.95 eV) for $\gamma$-phase indicates its suitability for optical and photocatalytic application. Yet,  most \cn-polymorphs suffer from structural instabilities \cite{dong2015,marques2004}, i.e., all these  have phonons with imaginary frequencies at zero applied pressure. {It is noteworthy that metastable $\gamma$-\cn~is already been experimentally observed by Andrade {\it et. al.} \cite{andrade2018}. The measurements found the emission signals for $\gamma$-\cn~, but signals disappear after few scanning tunneling microscopy (STM) scans. As the experiments  were performed at atmospheric conditions, where oxygen is always present and STM scans can generate high-energy photons, which possibly oxidizes the $\gamma$-\cn~phase. This indicates towards metastability of $\gamma$-\cn~at ambient conditions ({i.e., at room temperature and one atmospheric pressure}). Therefore, we investigate the pressure effect on $\gamma$-phase.} In Fig.~\ref{fig7}, we plot phonons at `zero' pressure (top) and at 275 GPa  (bottom). The 0 GPa phonons show imaginary frequencies that indicates a dynamical instability in $\gamma$-phase. However, under hydrostatic pressure (above 275 GPa) all imaginary frequencies in Fig.~\ref{fig7} disappear, i.e., above a critical pressure $\gamma$-phase become dynamically stable.  

%%% FIGURE 10

{\par}The dynamically stable $\gamma$-phase also satisfies the Born stability criteria \cite{PhysRevB.90.224104}, i.e., (a) $C_{11} - C_{12} > 0$;  (b) $C_{11} + 2C_{12} > 0$;   (c) $C_{44} > 0$, where  $C_{11} =436.5$~GPa, $C_{12} = 363.09$~GPa, and $C_{44} = 446.5$~GPa. The strong covalent bonding gives large bulk-, elastic-, and shear-modulus of 387.6 GPa, 682.0 GPa and 282.6 GPa, respectively.  

\begin{figure}[t]
\centering
\includegraphics[scale=0.4]{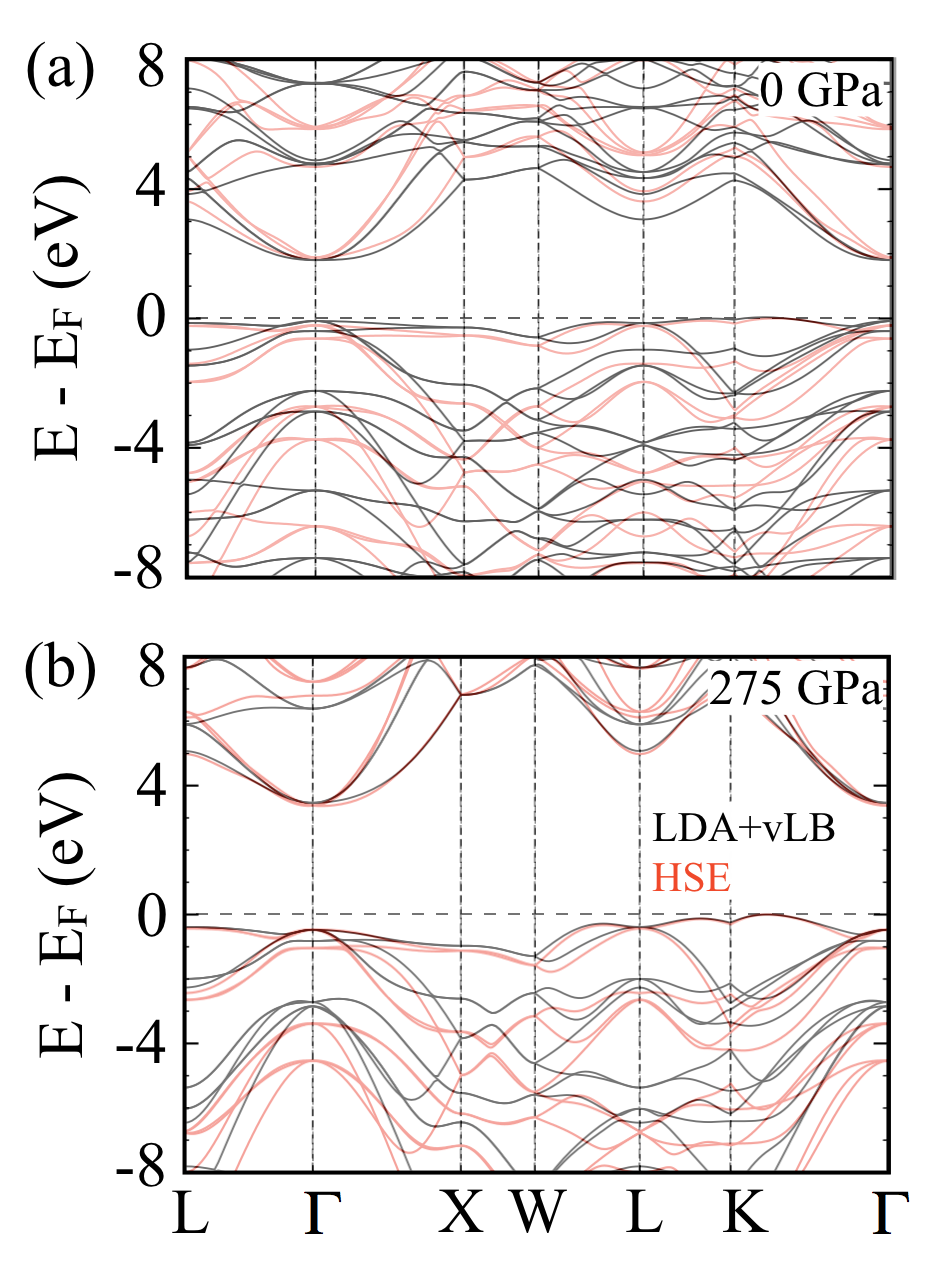}\\ 
\caption{{The LDA+vLB and QE-HSE calculated electronic band-structures of $\gamma$-\cn~at (a) 0 GPa, and (b) 275 GPa. The shape and position of conduction band minima and valence band maxima calculated using LDA+vLB and QE-HSE for $\gamma$-\cn~in (a) and (b) shows good agreement}.}
\label{fig12}
\end{figure}

{\par}{To point out the effect of pressure on band-structure, we plot electronic band-structures of $\gamma$-\cn~ in Fig.~\ref{fig12}(a)$\&$(b). For with and without pressure cases, $\gamma$-\cn~ shows good agreement for conduction band minima and valence band maxima positions, which is important both for calculation of optical properties and band alignment in determining photocatalytic behavior.}

\begin{figure}[t]
\centering
\includegraphics[scale=0.3]{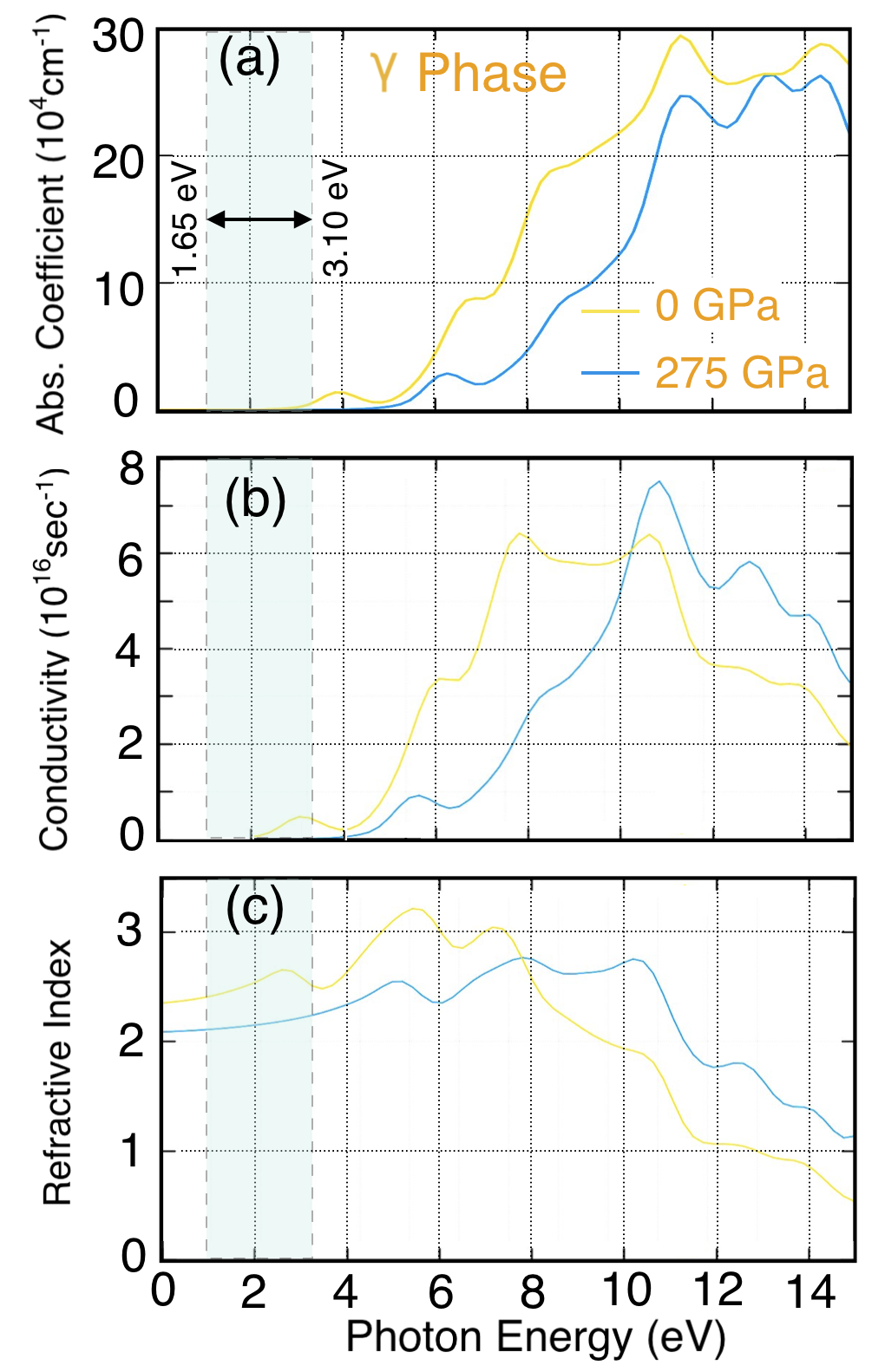} 
\caption{(a) Absorption spectra, (b)  optical conductivity, and (c) refractive index versus photon energy for $\gamma$-C$_{3}$N$_{4}$ at `zero' pressure (yellow) and 275 GPa (blue) calculated using HSE functional. The shaded zone indicates visible light range (1.65$-$3.10 eV).}
\label{fig8}
\end{figure}

{\par}The effect of pressure clearly reflects on optical properties of $\gamma$-\cn~ (Fig.~\ref{fig8}). We found an increase in the extent of absorption edge for the $\gamma$ phase compared to ambient ({one atmospheric}) pressure (Fig.~\ref{fig8}(a)). The absorption edges at $\sim$300~nm for $\gamma$ phase corresponds to the band gap energy of 3.43 eV. First, the optical absorption range for $\gamma$ phase increases under pressure, and, secondly, the optical conductivity of $\gamma$-\cn~  (Fig.~\ref{fig8}(b)) also changes from that at ``zero pressure'' (Fig.~\ref{fig5}(b)). A slight change in conductivity compared to ``zero-pressure'' suggests increased photocatalytic behavior of $\gamma$-phase. Refractive index of dynamically stable $\gamma$-phase in Fig.~\ref{fig8}(c) reduces compared to ``zero pressure'' case. The calculated optical gap corresponds to the photon energy at which the imaginary part of the refractive index, $k$, becomes non-zero. We find the refractive index for $\gamma$-\cn~changes 2.5 to 2.05 for the visible light range. 

\begin{figure}[H]
\centering
\includegraphics[scale=0.32]{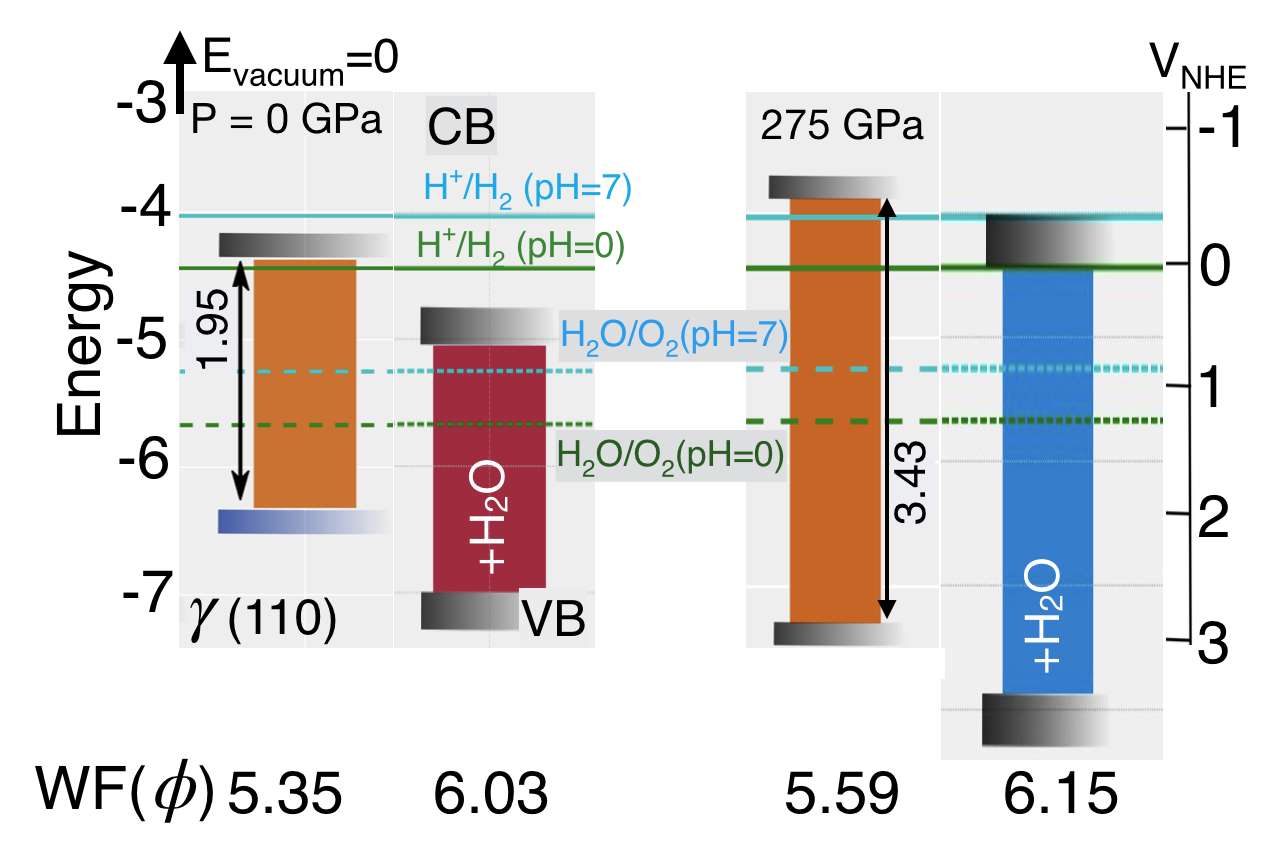} 
\caption{At $0$ GPa (dynamically unstable) and $275$ GPa (stable), the conduction and valence band positions of favorable (110) surface with and without water of $\gamma-C_{3}N_{4}$-polymorph with respect to water reduction and oxidation potentials.}
\label{fig9}
\end{figure}

{\par}In Fig.~\ref{fig9}, we compare the VB and CB positions and work function for most favorable surface (110) of $\gamma$ phase with(out) pressure. Clearly, the photocatalytic range of $\gamma$ (110)-surface under pressure (275 GPa) increases. We also showcase the effect of water on (110) surface on $\gamma$ phase and find slight reduction in CB band-edge compared to surface with no water. The CB band-edge becomes slightly less positive compared to pure (110) surface, whereas, the $\gamma$ (110) phase is now more favorable under pressure due to increased photocatalytic range (pH=0 to 7) in both with and without water case. 

	{The pressure profile for the structural stability of $\gamma$-\cn~is significantly higher, however, recent studies presented novel and more practical synthesis routes for new or known high-pressure phases under predictable nonhydrostatic loading \cite{PRL2018_DDJ}. It has been exemplified for silicon that the hydrostatic pressure of 76 GPa could be lowered by 21 times to 3.7 GPa under uniaxial loading. We believe that the similar idea could be employed in future for synthesizing new high-pressure carbon-nitride materials. To add further discussion on pressure and point out its importance, recently, pressure has helped discover new material(s) and mechanisms \cite{NatChem2017, Cabon2019}. Although, the stability of $\gamma$-\cn~is an issue \cite{Pradhan2010,PhysRevB.70.104114}, the use of hydrostatic pressure in this work is an effort to provide theoretical understanding of metastable (structurally unstable) $\gamma$-\cn.}

%%% FIGURE 11

\section{Conclusion}
We present a detailed study of both structural and electronic-structure of \cn~polymorphs using a localized-basis (FP-LMTO-LDA+vLB) and a plane-wave basis (QE-HSE) method. The structural properties and band gap of \cn-polymorphs predicted from LDA+vLB agrees well with our more accurate QE-HSE results, and available experiments. 
We did cross-validation tests performed on two diverse cases:  (a) 2D-graphene, and (b) bulk-Si. The calculated band gaps (graphene, 0 eV; bulk Si, 1.25 eV indirect) agree well with experiments, which shows the predictive reliability and accuracy. 
{To point out important differences between our approach and other calculations done for C$_{3}$N$_{4}$ in literature are two-fold, (a) most calculations are performed using LDA/GGA functionals, which are known to underestimate the band gap;  and (b)  hybrid  or GW functionals were not used in systematic way to study of relevant governing factors for photocatalytic performance of carbon-nitrides.} 

{\par} We performed optical-property, and work-function calculation using hybrid-functional (HSE) as implemented within Quantum-Espresso package.
Beside lattice parameters and band gaps, work-function calculations are found to be in very good agreement with the available experiments.
Our calculations show optimal band gap of 1.95 eV as well as higher optical conductivity and carrier mobility for $\gamma$-\cn~than other polymorphs, which makes $\gamma$-\cn~a suitable candidate for photocatalytic applications. The work function in $\gamma$-\cn~shows an orientation dependence with (110) surface as most favorable orientation over (100) and (111). In spite of favorable electronic-structure and optical properties, $\gamma$-phase shows a dynamical instability (unstable phonons) at standard temperature and pressure.  
We show that $\gamma$-\cn~is structurally stabilized under hydrostatic pressure (275 GPa) and leads to a significant improvement in photocatalytic behavior with respect to water reduction and oxidation potentials. We could also identify the active sites in \cn-polymorphs from charge-density plots as strongly localized states, usually suggesting low photo-absorption efficiencies. 
Our study offers $\gamma$ phase as a new candidate to explore for optical and photocatalytic applications, which may further ignite interest in \cn~polymorphs within the materials and chemistry community.

\section*{ACKNOWLEDGMENT}
SD and PS equally contributed. vLB routines for TB-LMTO-ASA and FP-NMTO are developed by PS/MKH/DDJ and SD/AM, respectively. Calculations are carried out by SD/DJ/CBC in collaboration with PS/DDJ. Research at Ames Laboratory was supported by the U.S. Department of Energy (DOE),  Office of Science, Basic Energy Sciences, Materials Science and Engineering Division. Ames Laboratory is operated for the U.S. DOE by Iowa State University under Contract No. DE-AC02-07CH11358. 

\biboptions{sort&compress}
\bibliography{Manuscript}
\bibliographystyle{rsc}

\end{document}